\documentclass[twocolumn,prb,preprintnumbers,amsmath,amssymb,floatfix]{revtex4}
\usepackage[linktocpage,bookmarksopen,bookmarksnumbered]{hyperref}
\usepackage{graphicx}
\usepackage{dcolumn}

\usepackage{amsmath,graphics,epsfig,color,verbatim,ulem}

\renewcommand{\vr}{{\mathbf{r}}}

\newcommand{\Tr}{\mathrm{Tr}}

\renewcommand{\Im}{\mathrm{Im}}

\begin{document}

\setlength{\pdfpageheight}{\paperheight}
\setlength{\pdfpagewidth}{\paperwidth}

\title{Covalency in transition metal oxides within all-electron Dynamical Mean Field Theory}
\author{Kristjan Haule}
 \email{haule@physics.rutgers.edu}
\author{Turan Birol}
\author{Gabriel Kotliar}
\affiliation{Department of Physics and Astronomy, Rutgers University, Piscataway, NJ 08854.}
\date{\today}
\begin{abstract}
A combination of dynamical mean field theory and density functional
theory, as implemented in Phys. Rev. B 81, 195107 (2010), is applied
to both the early and late transition metal oxides.  For a fixed value
of the local Coulomb repulsion, without fine tuning, we
obtain the main features of these series, such as the metallic
character of SrVO$_3$ and the the insulating gaps of LaVO$_3$,
LaTiO$_3$ and La$_2$CO$_4$ which are in good agreement with
experiment.  This study highlights the importance of local physics and
high energy hybridization in the screening of the Hubbard interaction
and how different low energy behaviors can emerge from the a unified
treatment of the transition metal series.
\end{abstract}
\maketitle

\section{Introduction}

The quantum mechanical description of electrons in solids -- the band
theory~\cite{Bethe,Sommerfeld,Bloch} --offered a straightforward
account for distinctions between insulators and metals. Fermi liquid
theory~\cite{AGD} has elucidated why interactions between
10$^{23}$~cm$^{-3}$ electrons in simple metals can be readily
neglected, thus validating inferences of free electron models. It came
as a considerable surprise in late 30s when crystals with incomplete
$d$ bands were found insulating~\cite{Mott1}.  The term "Mott
insulator" was later coined to identify a class of solids violating
the above fundamental expectations of band
theory~\cite{Mott3}. Peierls and Mott stated~\cite{Mott2} that "a
rather drastic modification of the present electron theory of metals
would be necessary in order to take these facts into account" and
proposed that such a modification must include Coulomb interactions
between the electrons. Study of correlations in solids, which are
responsible for such a dramatic increase of resistivity in Mott
insulators, remains in the forefront of contemporary condensed matter
physics~\cite{Imada,optics-rev}, and it was later found in many other
materials, such as $d$ and $f$-electron inter-metallic compounds, as
well as a number of $\pi$-electron organic conductors.

The theory became predictive with the invention of the Density
Functional Theory (DFT)~\cite{Kohn}. Within the Kohn-Sham framework,
the computation of the density of the solid is reduced to a tractable
problem of non-interacting electrons moving in an effective potential.
The implementation of DFT within the local density approximation (LDA)
and generalized gradient approximations (GGA) in 1970s, and the
increase in computational power in the past decades made it possible to
predict materials properties \textit{ab initio}. In weakly
correlated materials the computed Kohn-Sham spectra is a reasonable
description of the electronic spectra.  However, materials with
strong correlations, and in particular Mott insulators, are not
properly treated within these approximations.

It's long been recognized that electron correlations are mostly local
in space -- two widely separated electrons are unlikely to be
significantly correlated. Within LDA, the Kohn-Sham potential in each
point of space depends solely on the density at the same point,
hence LDA is local for each point in 3D-space.  However, in solids
with partially filled $d$-bands, the correlations are very strong
between two electrons on the same transition metal ion, which is
beyond the scope of LDA.

In 1990s the Dynamical Mean Field theory
(DMFT)~\cite{DMFT-first,Vollhardt,DMFT-RMP1996,PhysToday} was developed.  This
theory introduces non-locality in time, which is essential for
description of paramagnetic Mott insulators. This theory is also a
local theory, but it is local to a given site rather than a point in space.
DMFT successfully predicted the Mott-transition in the Hubbard
model~\cite{DMFTPhases,DMFTPhases2,GKrauth}, as well as many other known features
of correlated systems, such as the dynamical spectral weight
transfer~\cite{Gabi-optics}, the existence of a Mott end-point, and
the value of the critical exponents at this Mott
end-point~\cite{Georges-exponents}. The cluster-DMFT
studies~\cite{Trambley} show that this properties are genuine to the
frustrated Hubbard model.

Within DMFT, the functional that contains all local correlations is known
exactly, and can be calculated by solving an appropriate quantum
impurity model, but the computational cost when including many
interacting degrees of freedom on a given site increases
exponentially, while the hybridization with non-interacting states
does not increase the computational cost significantly.  At present,
modern computers allow us to treat interactions exactly within a
complete $d$ shell of a transition metal ion or a complete $f$ shell
in an inter-metallic compound, while the rest of the states must be
treated by a mean-field method. The most popular choice of such a
mean-field method is DFT, hence the combination of the two methods,
first proposed in 1997~\cite{Gabi-first}, was named
LDA+DMFT~\cite{review,Held}. The method became very successful as it
could predict properties of an extraordinary number of correlated
materials previously resisting detailed material specific predictions
(for a review see~\onlinecite{review,Held}).
The electronic structure and unusual physical properties of many
actinides~\cite{Savrasov-Nature,Savrasov-Science,shim2007},
lanthanides~\cite{zolfl2001, held2001,ylvisaker2009,GeorgesCe},
$3d$~\cite{pavarini2004,Pavarini_LaVO3,Pavarini,nekrasov2005,ohta2012},
$4d$~\cite{gorelov2010,RuOurs} and $5d$~\cite{Oursiridates} transition
metal compounds were explained using this approach.

In the early 2000s, the LDA+DMFT method was usually referring to the
dynamical mean field calculation of a lattice model, namely the
Hubbard model, where the hopping parameters were derived by a so-called
downfolding procedure: The bands near the Fermi level were
represented in terms of a small number of Wannier states\cite{WannierOrig}, and the
resulting Hubbard model was solved by the DMFT method. The feedback of
correlations to the electronic charge distribution, and hence
the Kohn-Sham potential, was often neglected. Also, usually the minimum number of
Wannier states were kept in the model, which made such model
calculations conceptually simple, but less predictive, as the Coulomb
repulsion for a low energy model is strongly screened by the degrees
of freedom eliminated from the model, hence a material specific and
model specific calculation of the interaction strength $U$ was needed
to make this method predictive.  The
constrained-RPA~\cite{constrained-RPA1,constrained-RPA2} was invented
for that purpose, and was quite successful when the correlations are
applied in the narrow energy window.

An alternative route, which avoids construction of the low-energy
model, was proposed by Savrasov \& Kotliar~\cite{Savrasov-PRB}, in
which the correction (self-energy) due to the correlations is added to the
Kohn-Sham potential in a very limited region of the real space, such
as the muffin-thin (MT) sphere of the correlated ion. In this approach, all
degrees of freedom local to an ion are treated exactly, while the
non-local correlations are treated in a mean-field way by DFT. No
valence state is eliminated from the model!  Kohn-Sham potential is
computed on the self-consistent electronic charge.  We call this
methodology the all-electron method.  The early implementation of this
approach, together with electronic charge self-consistency,
successfully predicted the phonon spectra of elemental
plutonium~\cite{Savrasov-Science}, but the impurity solvers at that
time were not adequate to address many other challenging problems in
correlated solids.
The DFT+DMFT method has rapidly matured over the last few years, as
several charge self-consistent implementations in various electronic
structure codes appeared~\cite{Haule-DMFT,Amadon1,Amadon2,Lecherman,Nordstrom,Aichorn},
some with integrated state of the art impurity
solvers~\cite{Aichorn,Haule-DMFT}.

The most significant difference between the earlier and more modern
implementations of the method is the degree of localization of
the electronic orbitals, which interact with strong Coulomb interaction.
In the early days, a set of Wannier orbitals spanning a
narrow window around the Fermi level was typically treated by the DMFT.
Since the non-local interactions and the non-local correlations are
neglected in the DMFT approach, one expects that a more localized choice
of orbitals leads to better results within single-site DMFT
approximation. Hence newer implementations applied correlations to
more localized states, and kept a larger number of itinerant states in
the model. A real space projector to the spherical harmonics within a
MT-sphere around the correlated ion $P_R(\vr\vr',lm
l'm')=Y_{lm}(\hat{\vr})\delta_R(r-r')Y^*_{lm}(\hat{\vr}')$ (where
$\delta_R(r-r')$ is nonzero when $r<R$ and $r'<R$ ) is a good example
of extremely localized orbitals, which hybridizes with a large number
of Kohn-Sham states, spanning a large energy window in band
representation. Such a set of real-space orbitals is clearly more
localized than the popular choice of maximally localized Wannier
orbitals\cite{WannierRMP}, which are constrained to faithfully
represent some set of low energy bands.

Numerous successful predictions of this all-electron DFT+DMFT were
published in the past decade nevertheless, its predictive power for
transition metal oxides was questioned recently in
Refs.~\onlinecite{Chris1,Chris2}.
Namely, using Wannier functions for oxygen-$p$ states and transition metal $d$
states, the authors of Refs.~\onlinecite{Chris1,Chris2} concluded that
fine tuning of several parameters, including the double-counting and
the interaction $U$ is needed to describe the Mott insulating state in  early
and late transiton metal oxides.
Moreover, the $p-d$ model requires
occupancy of the $d$ orbitals to be close to unity for the Mott state,
while DFT solution projected to the orbitals of their choice, predicts
far larger occupancies, hence this discrepancy
between DFT occupancies and the DMFT requirements lead them to
suggest that the self-consistent DFT+DMFT can not describe the Mott
insulating state without fine tuning the interaction $U$ to be in
the narrow range of $6\pm 1\,$eV and ad-hoc fine tuning of the
double-counting to reproduce 
the experimentally observed $p-d$ splittings.
This calls for a critical reexamination of  the application of the LDA+DMFT to the 3$d$ series.

Our methodology~\cite{Haule-DMFT} was tested in numerous classes of
materials, such as
actinides~\cite{Ours_Pu,Ours_Pu2,BranchingRatios,Ours_URu2Si2,Ours_Pu3,NeutronFF,NuclearFuels,Zhu_1,PuSb},
lanthanides~\cite{shim2007,Haule-DMFT,Shim_1,Shim_3}, transition metal
oxides~\cite{LaNiO3,LaNiO3_2,FeO}, iron
superconductors~\cite{Zhip_NatP,Zhip_NatM,Pengcheng_1,Shim_2}, and
other transition metal compounds~\cite{Oursiridates,FeSi}.
%
However, results
for early and late transition metal oxides with our methodology are not available in
literature.
%
It is therefore important to test our methodology in this
class of materials, which have mostly been studied using downfolded
LDA+DMFT implementations.

\section{Method}
\label{Method}

In this work we perform DFT+DMFT calculations for a series
of early transition metal oxides: SrVO$_3$, LaVO$_3$, LaTiO$_3$, and
a cuprate parent compound La$_2$CuO$_4$; all the test cases which required
fine tuning in Refs.~\onlinecite{Chris1,Chris2}.
We show that no fine tuning or adjustable parameter is required in
DFT+DMFT implementation of Ref.~\onlinecite{Haule-DMFT}, and for a fixed
value of on-site Coulomb repulsion $U=10\,$eV Mott gaps in all these
compounds are in reasonable agreement with experiment.

The all-electron DFT+DMFT
implementation~\cite{Haule-DMFT} extremizes the following
functional~\cite{review} 
\begin{widetext}
\begin{eqnarray}
\Gamma[\rho,V_{KS},G_{loc},\Sigma,V_{dc},n_d]=
-\Tr\ln\left(
(i\omega+\mu+\nabla^2-V_{KS})\delta(\vr-\vr')-\sum_{\tau LL'}P(\vr\vr',\tau LL')(\Sigma-V_{dc})_{L'L} \right)\\
-\int [V_{KS}-V_{ext}]\rho d^3r
-\Tr\left(\Sigma G_{loc}\right) + \Tr(V_{dc} n_d)+
\Phi_H[\rho]+\Phi_{xc}[\rho]+\Phi_{DMFT}[G_{loc}]-\Phi_{dc}[n_d]\nonumber
\end{eqnarray}
\end{widetext}
of three pairs of conjugate variables. At the saddle point,
$\rho$, $V_{KS}$ are the electronic charge
density, the Kohn-Sham potential, 
$G_{loc}$
and $\Sigma$ are the local Green's function and DMFT self-energy,
$V_{dc}$ is the double-counting potential, and $n_d$ is the occupancy
of the correlated orbital. $\Phi_H[\rho]$ and
$\Phi_{xc}[\rho]$ are the Hartree and the exchange-correlation energy
functionals, $\Phi_{DMFT}[G_{loc}]$ is the sum of all skeleton
diagrams constructed from $G_{loc}$ and local Coulomb repulsion
$\hat{U}$. This summation is carried out by the impurity solver.
The local Coulomb repulsion $\hat{U}$ is parametrized with
the Slater parametrization with $J_{Hunds}=0.7\,$eV, and, if not otherwise
stated, $U=10\,$eV.
The impurity model is solved by the continuous-time quantum Monte
Carlo method~\cite{Haule-ctqmc,Werner}.

  $V_{ext}$ is the external potential, containing the
  material specific information.  $P(\vr\vr',\tau LL')$ is the
  projector to the local correlated orbital at atom $\tau$ with
  angular momentum indices $L$,$L'$. We use projector
  $P^{2}(\vr\vr',\tau LL')$ introduced in Ref.~\onlinecite{Haule-DMFT}
  with an energy window of $\approx$20$\,eV$ around the Fermi
  level. For the maximal locality of correlated states, this projector
  is implemented in real space and is nonzero only within the
  MT-sphere of the correlated ion. In
  section~\ref{Comparison} we test several different projectors,
  ranging from extremely localized to moderately delocalized, to
  understand the controversy in the literature regarding the DFT+DMFT
  results for the transition metal oxides.
  The vanadium and titanium $t_{2g}$ states are treated dynamically,
  while the empty $e_g$ states are treated by a static mean field. The
  copper ion with its almost full shell requires dynamic treatment of
  all five $3d$ orbitals.

For the double-counting correction, we used two methodologies: a) the
fully-localized-limit (FLL) formula introduced in Ref.~\onlinecite{Anisimov}
is used in section~\ref{Comparison} to ensure that the results are
robust and that the simplification used elsewhere does not change the
results appreciably from this standard prescription. b) the method
explained in Ref.~\onlinecite{Haule-DMFT} is used in most of this
paper, where $\Phi_{dc}[n_d]= n_d V_{dc}$ and $V_{dc}$ is also
parameterized by the standard fully-localized-limit
formula~\cite{Anisimov} $V_{dc}=U(n_d^0-1/2)-J/2(n_d^0-1)$, and
$n_d^0$ is taken to be the nominal occupancy of the correlated ion. We
name this method fixed-DC. In particular, for SrVO$_3$ with V$^{4+}$
ion we take $n_d^0=1$, for LaVO$_3$ with V$^{3+}$ ion $n_d^0=2$, for
LaTiO$_3$ with Ti$^{3+}$ ion $n_d^0=1$ and for La$_2$CuO$_4$ with
Cu$^{2+}$ ion $n_d^0=9$. This double-counting scheme has two virtues:
i) it is numerically much more stable in the charge-self-consistent
DFT+DMFT, as the noise from Monte-Carlo does not feed-back into
impurity levels, and into large Hartree shifts. ii) the results are
more robust with respect to small changes in projector, linearization
energies, etc.
Both double-countings are equally justifiable on the phenomenological
level. The determination of the exact double-counting is an open
problem, but see recent progress in Ref.~\onlinecite{exact-DC}.

The double-counting b) ensures that at infinite $U$ one recovers
atomic physics at the nominal valence. For discussion's sake, let us set
$J_{Hunds}$ to zero. In the absence of any double-counting correction,
the lower Hubbard band in the atomic limit is positioned at
$\varepsilon_f+U (n_d^0-1)$, and the upper Hubbard band at
$\varepsilon_f+U n_d^0$, where $\varepsilon_f$ is the center of the
correlated state at $U=0$ (in DFT calculation). The center between the
Hubbard bands is at $\varepsilon_f+U(n_d^0-1/2)$. To ensure that in
the large $U$ limit the center of the correlated states does not move
from its DFT position, and that we recover the correct nominal
occupancy, we must subtract from the dynamic self-energy the
correction $U(n_d^0-1/2)$, which brings the center of the correlated
state to its center in DFT. Hence a good choice for double counting
correction (in the absence of Hunds coupling) is given by
$U(n_d^0-1/2)$ with $n_d^0$ the nominal valence.  In typical model
calculations for the downfolded Hubbard model, such nominal valence is
automatically enforced.

The DFT part of our code is based on the WIEN2k
package~\cite{wien}. The exchange-correlation energy in DFT ($\Phi_{xc}[\rho]$) is
evaluated using the PBE functional.\cite{PBE}
The DFT+DMFT calculations are fully self-consistent in the electronic
charge density, chemical potential, and impurity levels.
The temperature is set to 200K. The experimental crystal structures
from Refs.~\onlinecite{SVO-struct},~\onlinecite{LVO-struct},~\onlinecite{LTO-struct}~\onlinecite{LSCO-struct}
are used for SrVO$_3$, LaVO$_3$, LaTiO$_3$, La$_2$CuO$_4$,
respectively.  To obtain spectra on the real axis, maximum entropy
method is used for analytical continuation of the
self-energy.\cite{max-entropy} Finally, the VESTA software is used at
various points to visualize and study the crystal
structures.\cite{VESTA}

\section{Results}
\label{Results}

\begin{figure}[h]
\centering{
\includegraphics[width=1\linewidth]{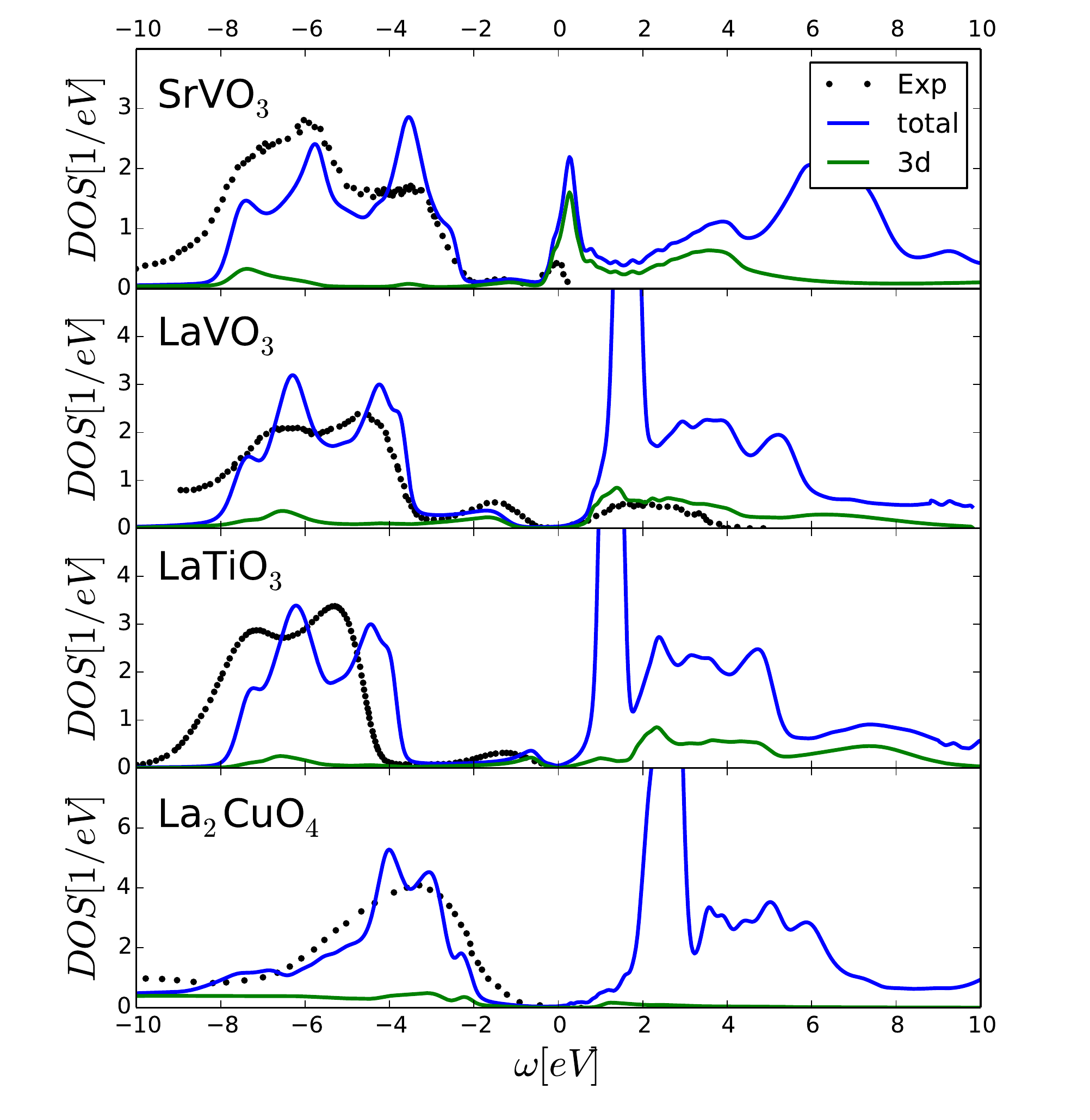}
  }
\caption{
  The total DOS and its projection to the $3d$ ion for
  selected transition metal oxides. Experimental photoemission for
  SrVO$_3$, LaVO$_3$, LaTiO$_3$, and La$_2$CuO$_4$ is plotted by black
  dots, and was digitized from
  Refs.~\onlinecite{SVO-PES},~\onlinecite{LVO-PES},~\onlinecite{LTO-PES},~\onlinecite{LSCO-PES}, respectively.
}
\label{DOSall}
\end{figure}
\begin{figure}[h]
\centering{
\includegraphics[width=1\linewidth]{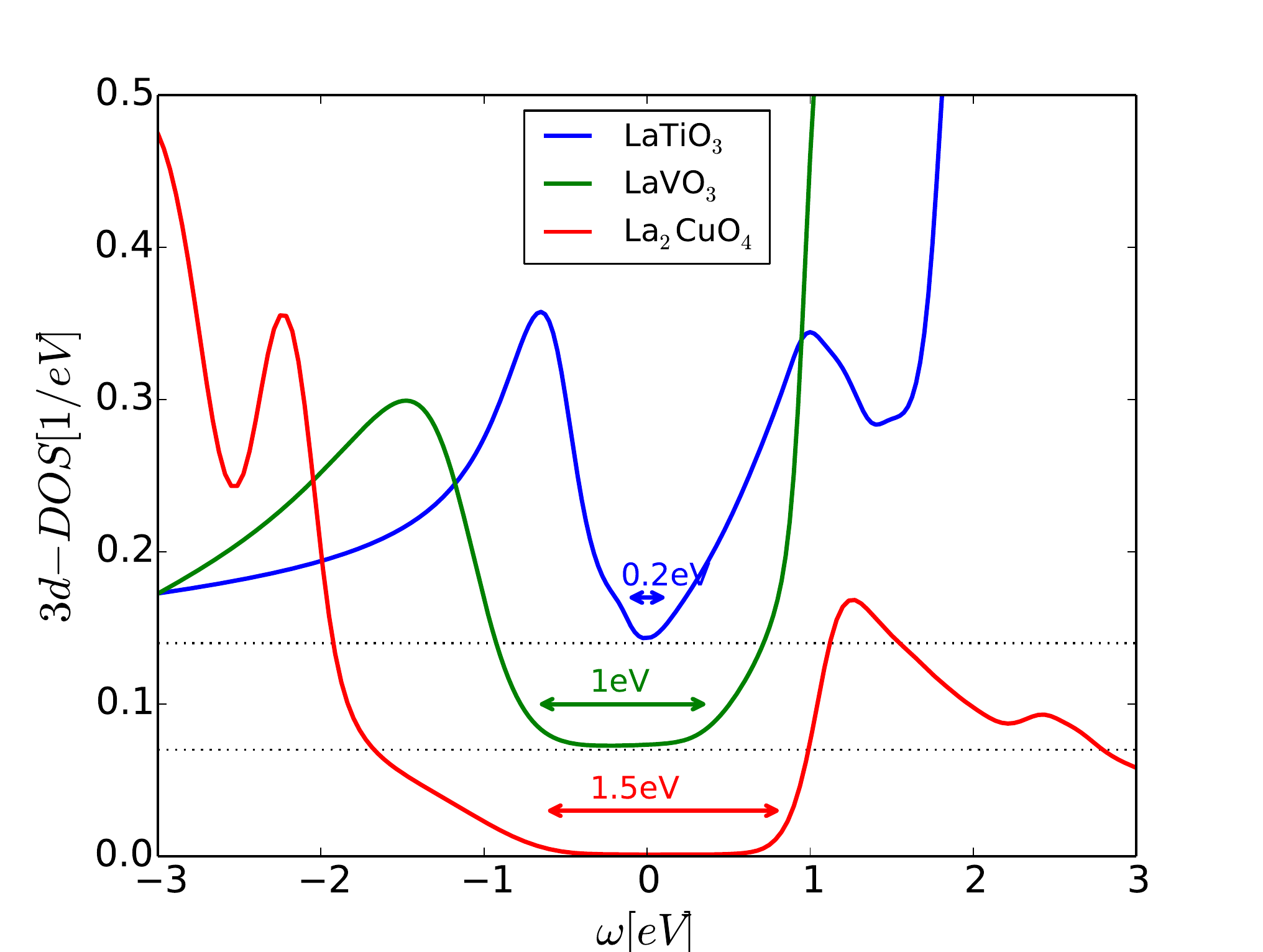}
  }
\caption{
  Zoom-in of the low-energy DOS projected to the $3d$ orbitals for
  insulating compounds. For clarity, the curves were offset for 0.07/eV.
  The arrows mark the experimental size of the
  gap.
}
\label{DOSgaps}
\end{figure}
Fig.~\ref{DOSall} shows the DFT+DMFT total and projected $3d$ densities
of states for the four test compounds. The photoemission measurements are
also shown for comparison. Fig.~\ref{DOSgaps} zooms the low energy
part of the DOS to display the gap sizes.  SrVO$_3$ is a mixed-valent
($n_d=1.19$) metallic compound with oxygen states centered around
-5 eV, a small shoulder corresponding to an incoherent excitation
(Hubbard band) around -1.5 eV of mostly $d$ character, and the quite broad
quasiparticle peak at the Fermi level with its bandwidth reduced from
DFT for roughly a factor of 2. These are all in excellent agreement with
the experiment~\cite{SVO-PES,SVO-MASS,SrVO3_ARPES}. Previous LDA+DMFT calculations of
Refs.~\onlinecite{pavarini2004,SrVO3_ARPES}, where only the $t_{2g}$ states were treated
in the model, gave very similar electronic spectra, hence results are
very robust with respect to the choice of the correlated
orbital. Notice that the value of $U$ depends on the choice of energy
window. Calculations with an energy window, which include only the
$t_{2g}$ states, requires a value of $U\approx 5\,$eV, as used in
Ref.~\onlinecite{pavarini2004}. For large energy window ($20\,$eV used
here) somewhat larger value of $U$ is needed, however, results are
reasonable for an extended range of $U$ values between 6-10\,eV.

LaVO$_3$ is a Mott-insulator with a gap size of approximately 1$~$eV
(see Fig.~\ref{DOSgaps}), in agreement with
experiment~\cite{LVO-PES}. The lower Hubbard band at -1.5$\,$eV, has a
considerably more admixure of oxygen-$p$ than SrVO$_3$, as noticed in
Ref.~\onlinecite{LVO-PES}.
LaTiO$_3$ has a very small Mott gap around 0.2 eV, similar to the
experimental gap~\cite{LTO-PES}. The Hubbard band is located at
$\approx -0.8\,$eV, and the oxygen-$p$ band edge is at $-4\,$eV.
Experimentally, the Hubbard band and oxygen band are located at
somewhat lower energy than theoretically predicted. Our DFT+DMFT
calculation does not shift the oxygen states appreciably from its DFT
position.
Finally, La$_2$CuO$_4$ is a wide-gap Mott insulator of charge-transfer
nature, and has a gap size of the order of 1.5$\,$eV and the position
of the oxygen-$p$ band around -3.5$\,$eV. The oxygen position is well
predicted by the theory, and also the gap value is in good agreement
with experiment~\cite{Imada}. Overall agreement with the experiment is
very satisfactory, considering that no tuning parameter is used in
these calculations.

To show that the fine tuning of local Coulomb repulsion $U$, which gets
screened by the valence states included in our DMFT calculations, is
not needed to get reasonable agreement with experiment, we show below
calculations for several values of $U$. We also display how valence changes 
with the increasing correlation strength $U$, and as
expected, we show that an infinite $U$ would lead to integer valence.
Notice that the DFT+DMFT valence in the actinides~\cite{BranchingRatios} agrees with
the experiment better than the LDA valence.

\begin{figure}[h]
\centering{
\includegraphics[width=1\linewidth]{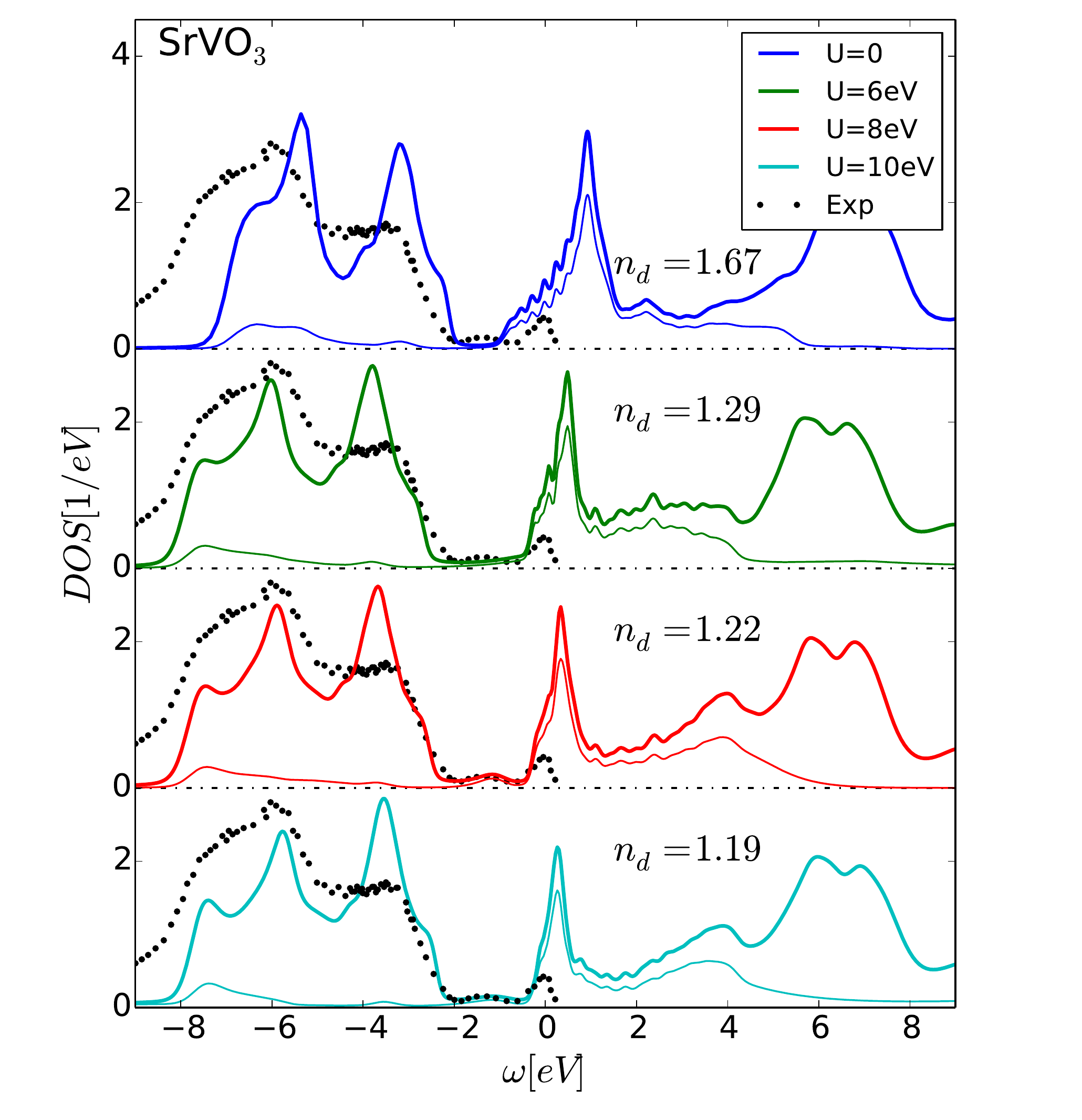}
  }
\caption{ The dependence of the electronic structure of SrVO$_3$ on
  the value of the local Coulomb repulsion $U$ in this all-electron
  calculation. The text displays the valence of $V^{4+}$ ion. The
  thick-line presents the total DOS and the thin-line its projection
  to the $3d$-ion subset.  For clarity, the different electronic
  structures are offset for 3eV.  }
\label{SVO}
\end{figure}
Fig.~\ref{SVO} shows the dependence of DOS in SrVO$_3$ on the local
Coulomb repulsion $U$. In the plot we also show the occupancy of the
V-$t_{2g}$ states as well as the photoemission spectra of thin
film~\cite{SVO-PES}. The $U=0$ results correspond to the GGA
calculation. The oxygen-$p$ bands move to a slightly ($<0.5\,$eV) lower
energy at $U=6$ eV while they progressively move back to its DFT
location at larger $U$.
The quasiparticle peak slightly narrows with increasing
$U$, but the Mott gap does not open even
for very large $U$ beyond 12 eV (not shown in the figure).
%
Notice that this is inconsistent with the universal phase diagram in Fig. 2 of
Ref.~\onlinecite{Chris2} since $n=1.2$ falls deep inside the
insulating regime for $d^1$ systems in that phase diagram.
The lower Hubbard band, located between the
quasiparticle peak and oxygen bands, appears at $U=8$ eV and becomes
even more pronounced at $U=10$ eV. The system is very mixed-valent
(non-integer occupancy) in GGA, but it becomes less mixed-valent with
increasing $U$. As expected, increasing correlation strength typically
reduces mixed-valency.
As is clear from Fig.~\ref{SVO}, results are not very
sensitive to the strength of the Coulomb repulsion, and any value of
$U$ between 6-12$\,$eV gives reasonable agreement with experiment. The
low energy $t_{2g}$ spectra is in good agreement with earlier DMFT
calculation of Ref.~\onlinecite{pavarini2004}, which included only the
$t_{2g}$ states.

\begin{figure}[h]
\centering{
\includegraphics[width=1\linewidth]{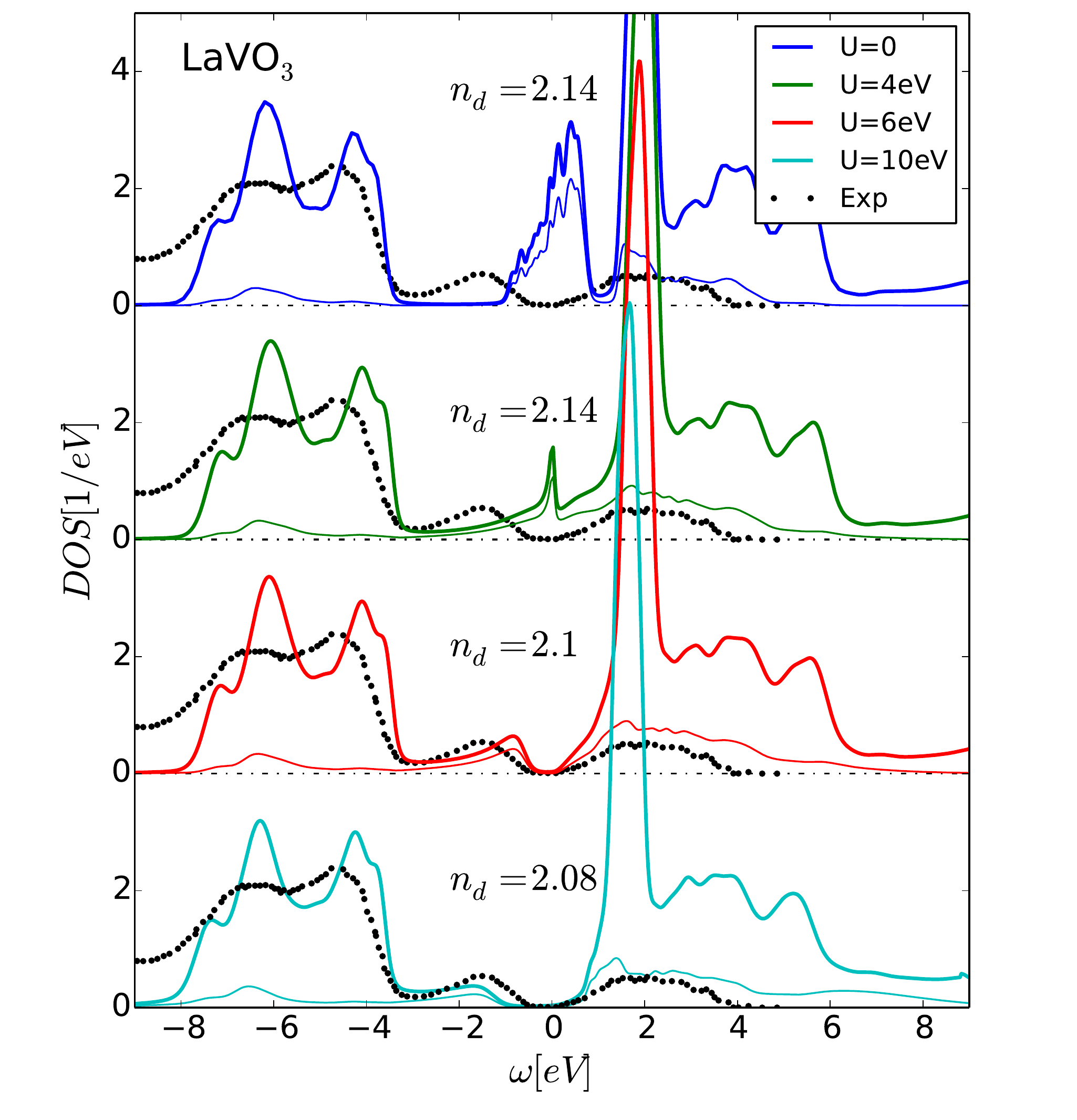}
  }
\caption{
The electronic structure of LaVO$_3$ for different values of
the local Coulomb repulsion $U$. We also display the valence of
$V^{3+}$ ion.
}
\label{LVO}
\end{figure}
Fig.~\ref{LVO} displays DOS for LaVO$_3$ for a range of $U$ values. In
GGA, the $t_{2g}$ states cross the Fermi level, while the oxygen-$p$ bands are
below -3.5$\,$eV and the La-$f$ bands are just above the $E_F$. For the correlation strength
of 4$\,$eV, LaVO$_3$ is still metallic, although the quasiparticle peak
becomes extremely narrow, while the valence does not change
appreciably from its GGA value. Larger $U=6\,$eV opens the Mott gap,
(the critical $U$ is approximately 5$\,$eV) and an
incoherent shoulder appears around $-1\,$eV, and mixed-valency is
reduced.  Finally at $U=10\,$eV the Mott gap is of the order of
$1\,$eV and the position of the incoherent shoulder is at $-1.5\,$eV
in good agreement with experiment~\cite{LVO-PES}. The partial
occupancies of the $t_{2g}$ orbitals at $U=10\,$eV are $n_{t2g}^1=0.56$,
$n_{t2g}^2=0.76$, and $n_{t2g}^3=0.76$, hence the orbital fluctuations are
strong due to the important role of Hund's coupling, which prevents
orbital polarization. This is all in good agreement with earlier DMFT
calculation of Ref.~\onlinecite{Pavarini_LaVO3}, which considered only the low
energy $t_{2g}$ degrees of freedom.
Notice that the oxygen-$p$ bands do not shift appreciably with
increasing $U$ and their position is well determined by DFT, in strong
contrast to the calculations in Ref.~\onlinecite{Chris1}, where the oxygen-$p$
bands move substantially for a small change in $U$.

\begin{figure}[h]
\centering{
\includegraphics[width=1\linewidth]{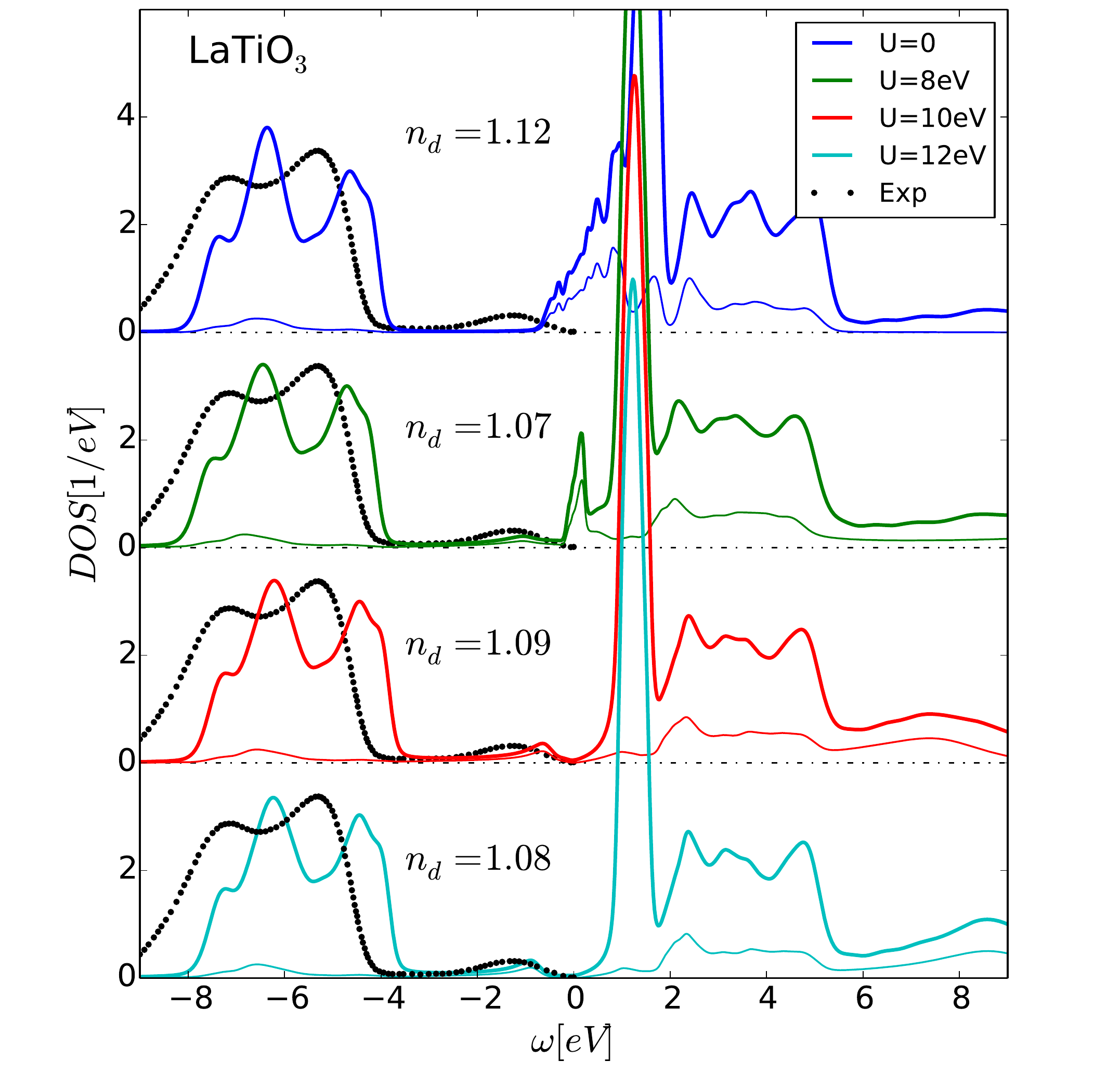}
  }
\caption{
The electronic structure of LaTiO$_3$ for different values of
the local Coulomb repulsion $U$. We also display the valence of
$Ti^{3+}$ ion.
}
\label{LTO}
\end{figure}
In LaTiO$_3$, displayed in Fig.~\ref{LTO}, the oxygen-$p$ bands start
around $-4\,$eV and a very sharp La-$f$ state appears just above the
Fermi level at 1$\,$eV. The position of this level is very sensitive
to the rotation angle of the oxygen octahedra, and a larger rotation angle,
which is obtained by the GGA-optimization of the structure, shifts the
level up by 0.5$\,$eV compared to experimentally determined crystal
structure~\cite{LTO-struct} used here.
This can be explained by the fact that octahedral rotations change the
coordination of the A-site cation (La) drastically.\cite{Woodward}
It is due to screening by this sharp La-$f$ state that the Mott
gap does not open for $U=8\,$eV and is very small ($\approx 0.2\,$eV)
at $U=10\,$eV. 

The orbital polarization is also very sensitive to the octahedral
rotation, as pointed out in Ref.~\onlinecite{pavarini2004}. For
the structure of Ref.~\onlinecite{LTO-struct}, the orbital polarization is modest in
our calculations ($n_{t2g}^1\approx 0.67$, $n_{t2g}^2\approx 0.20$,
$n_{t2g}^3\approx 0.20$) but for only slightly larger rotation, as
measured in Ref.~\onlinecite{LaTiO3_struct2}, the polarization is
almost complete ($n_{t2g}^1\approx 0.93$, $n_{t2g}^2\approx 0.06$,
$n_{t2g}^3\approx 0.08$).

Due to the substantial hybridization of the $t_{2g}$ states with the La-$f$
states very near the Fermi energy (see Fig.~\ref{Hyb}), the screening
is much stronger than it would be in the absence of this La-$f$ level.
Models which remove the La-$f$ states from consideration would hence need
substantially reduced interaction $U$ to reproduce the experimentally
observed Mott gap~\cite{LTO-PES}. Indeed, our results clearly
disagree with the phase diagram of Ref.~\onlinecite{Chris2}, which
leads us to believe that the $U-n_d$ phase diagram is more material
specific, and also depends on presence of other states that can screen
the interaction.
%
To open a Mott gap, a near integer occupancy is needed, but the
critical value of $U$ and $n_d$ can be quite material specific. Notice
that the double counting is also proportional to the Coulomb
interaction $U$, hence in the large $U$ limit, the upper and lower
Hubbard bands are pushed to positive and negative infinity
respectively; always resulting in an integer occupancy.
%
We also display photoemission measurements for
La$_{1-x}$Sr$_x$TiO$_{3+y}$ with $x+y=0.04$ from
Ref.~\onlinecite{LTO-PES}.  The oxygen-$p$ bands' position is somewhat
different than in experiment. However, this disagreement between DFT
and experiment might partially be attributed to slightly different
chemical composition of the crystal.  The local 
correlations on Ti do not improve the position of O-$p$ state.

\begin{figure}[h]
\centering{
\includegraphics[width=0.8\linewidth]{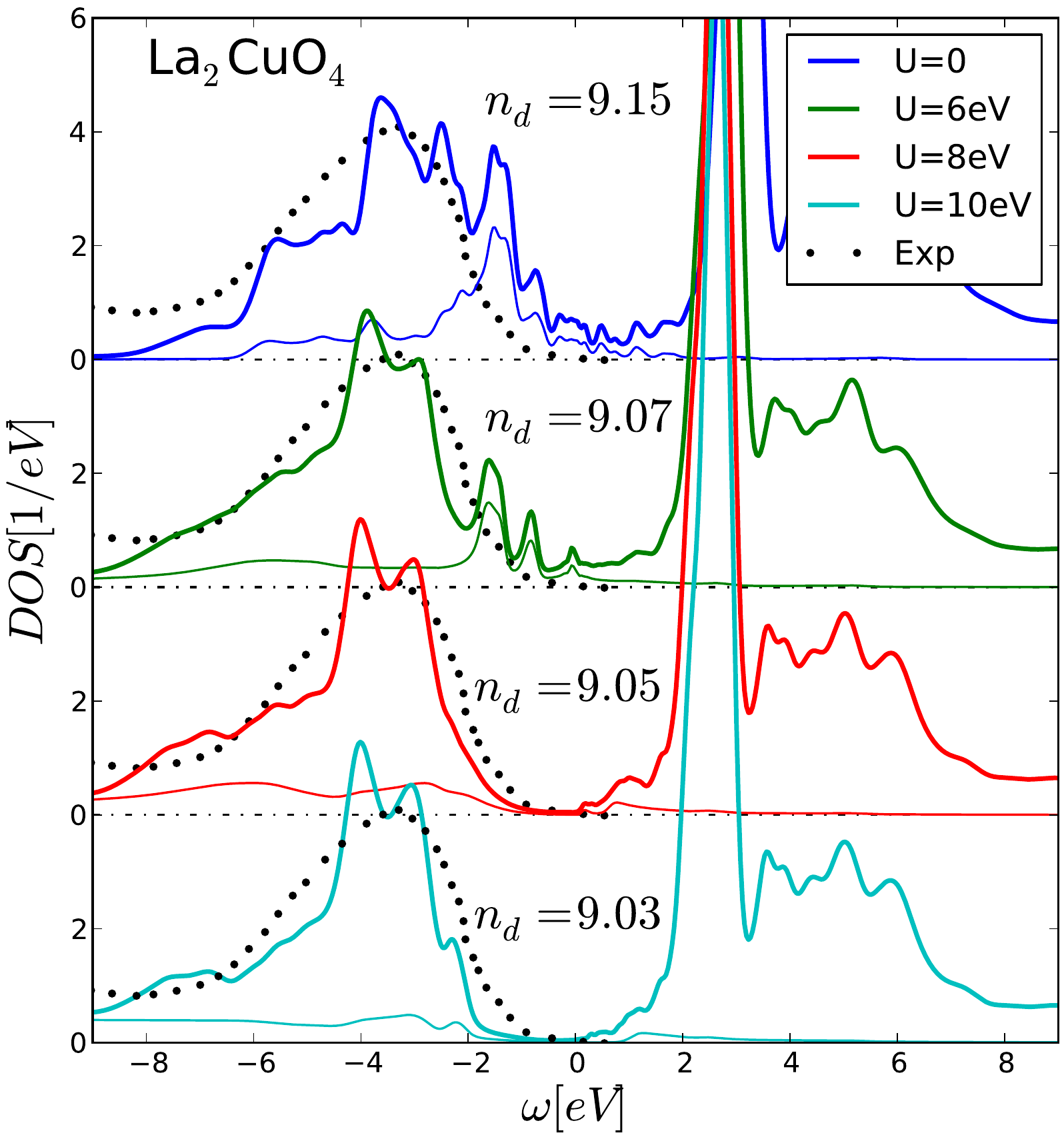}
  }
\caption{
The electronic structure of La$_2$CuO$_4$ for different values of
the local Coulomb repulsion $U$. We also display the valence of
$Cu^{2+}$ ion.
}
\label{LCO}
\end{figure}

\begin{figure}[h]
\centering{
\includegraphics[width=1\linewidth]{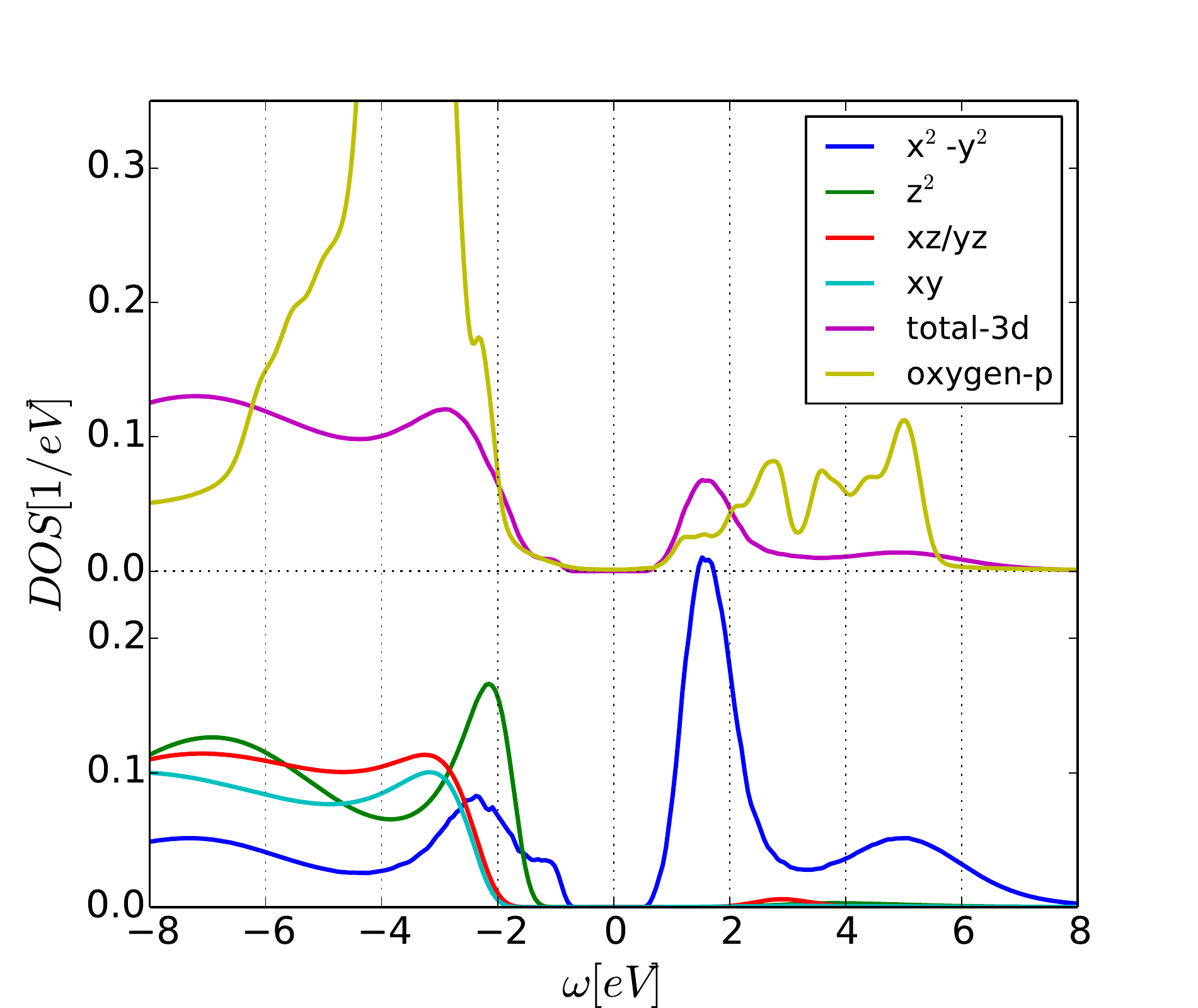}
  }
\caption{
Orbitally resolved DOS for La$_2$CuO$_4$ at $U=10\,$eV. The total-$3d$ and total
oxygen-$p$ DOS is offset for 0.25/eV.
}
\label{LCO2}
\end{figure}
Fig.~\ref{LCO} shows the DOS for La$_2$CuO$_4$, a late transition metal
oxide, and the parent compound of cuprate superconductors.  Within GGA 
($U=0$), the $e_{g}$ states cross the Fermi level and the
$t_{2g}$ states start at -1$\,$eV and strongly overlap with oxygen,
which starts at -2$\,$eV. The sharp peak at 2.5$\,$eV is due to the
La-$f$ states. The screening by La-$f$ states in cuprates is much weaker
than in LaTiO$_3$, as the hybridization function, displayed in
Fig.~\ref{Hyb}, has only a very weak peak at higher energy (2.4$\,$eV)
which couples primarily to $t_{2g}$ states, but not to low energy
$x^2-y^2$ orbital.  Once again, the oxygen-$p$ states and the La-$f$
states do not move appreciately with increasing correlation
strength. The Mott gap opens around $U=8\,$eV. As in the model
Hamiltonian studies within single site DMFT, 
once the insulating phase is reached~\cite{DMFTPhases}, the
gap opens discontinuously, and is of the order of $1\,$eV.  At
$U=10\,$eV, it is approximately 1.5 eV, in agreement with
the experiment~\cite{Imada}.  The oxygen bands' position is also in a good
agreement with the experiment~\cite{LSCO-PES}, centered at -3.5 eV.
Fig.~\ref{LCO2} resolves DOS in orbital space. Clearly, the $x^2-y^2$
orbital is half-filled, and in the interval between -1 eV and -2 eV the
amount of total-$3d$ states and oxygen states is almost exactly
equal. This is the region of the Zhang-Rice singlet. The $z^2$ orbital
is sharply peaked slightly below -2 eV, and the $t_{2g}$ states start
at -3 eV.

\begin{figure}[h]
\centering{
\includegraphics[width=1\linewidth]{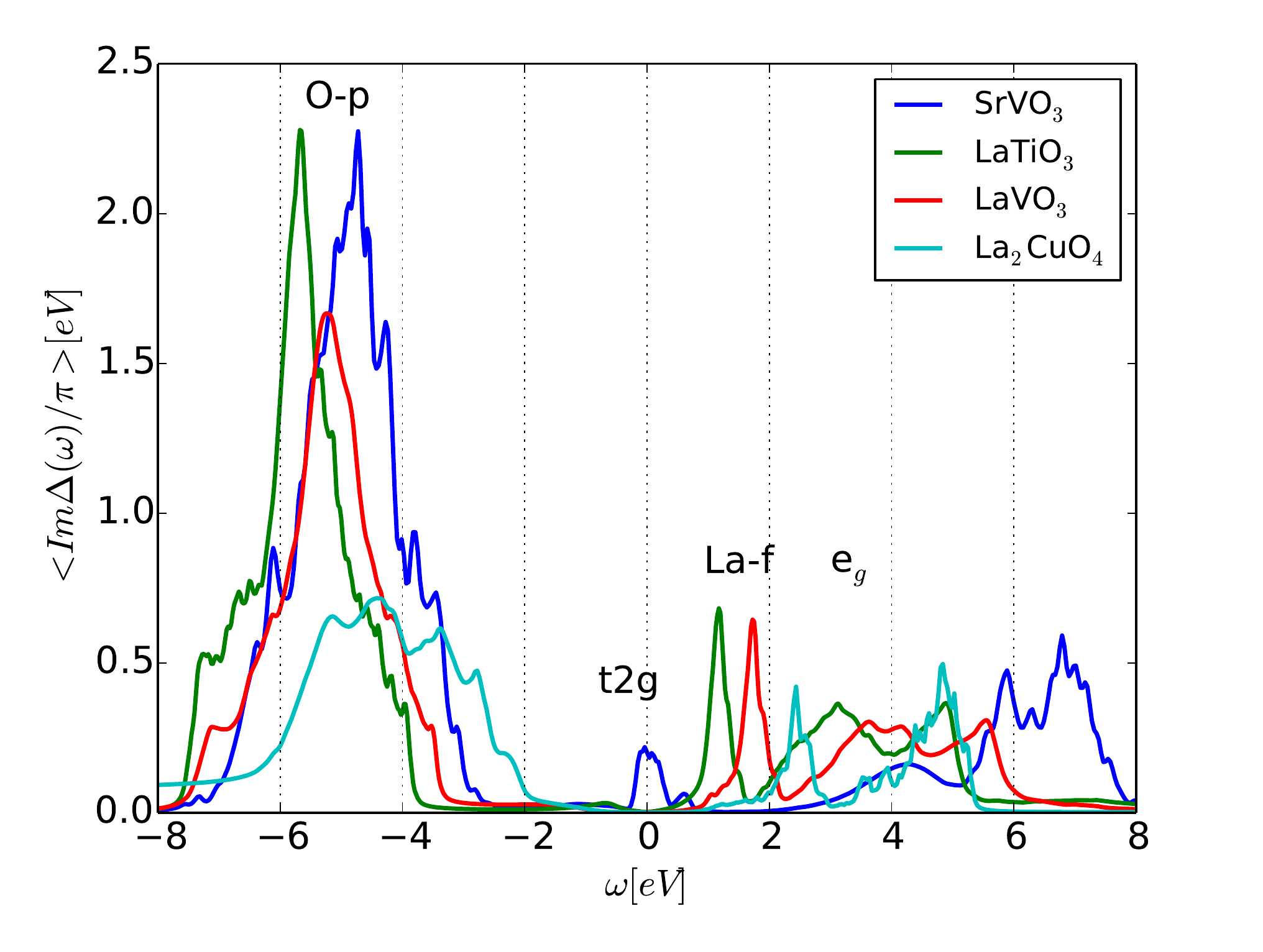}
  }
\caption{
The imaginary part of the impurity hybridization function for various
materials at $U=10\,$eV. We also display what is the dominant source of various
peaks in the hybridization function.
}
\label{Hyb}
\end{figure}

A main finding of this paper is that the all-electron treatment with a large energy
window requires a much larger value of $U$ than the corresponding
LDA+DMFT study with the small energy window, which keeps the $t_{2g}$ (or
$e_g$) states only.
The bare Coulomb repulsion $U$ of a model where all valence
states are kept in the model is screened only by the degrees of
freedom eliminated from the model, which are mostly core and semicore
states, resulting in a much larger on-site Coulomb repulsion $U$ in
this treatment, but also more universal $U$ across similar
compounds. The screening from model-$U$ down to the fully screened
interaction $W$ is achieved here by the impurity solver (see related
discussion in the context of the model in Ref.\onlinecite{Sordi2}).

On the one particle level, it is best demonstrated by the impurity
hybridization function, which contains signatures of all the valence
states in the solid that screen the local degrees of freedom.
We show in Fig.~\ref{Hyb} the average hybridization function (average
over orbitals) for an electron on transition metal site in different
compounds.  We can understand qualitatively the influence of the
window on the characteristic scale of the impurity model from the
following argument.  The Kondo scale should not depend on the energy
window chosen in the calculation. The Kondo scale is a function of
$U$ and impurity hybridization $T_k\approx exp(-const*U/\langle
Im\Delta\rangle_0)$, where $\langle Im\Delta\rangle_0$ is some
weighted average of the displayed hybridization function, with larger
weight given to the low energy. When only the $t_{2g}$ states in early
transition metal compounds are kept in the model, $U$ has to be small,
because the hybridization function is nonzero only within the narrow
region around $E_F$, and its average value is below 0.2$\,$eV.  When
oxygen-$p$ states are added, $\Im\Delta$ increases tremendously in the
interval $-7\,$eV to $-3\,$eV, and so does the weighted average
$\langle Im \Delta\rangle_0$, hence $U$ must increase to keep low
energy scale intact. It is also clear from Fig.~\ref{Hyb} that the La-$f$
states screen the Coulomb repulsion rather well, since they are
located very near the Fermi level; hence elimination of La-$f$ states
from the model would need to be compensated by a very material
specific $U$. Finally, some extra screening is also coming from $e_g$
states above the Fermi level, and since their position is very
material dependent, their elimination requires further tuning of
$U$. However, when all these states are kept in the model, the large
part of the screening is already contained in the model, and hence
the all-electron $U$ is rather large and more universal.


\section{Discussion and Comparison to Other Studies}
\label{Comparison}

To asses the robustness of our results, we also tested the widely used
fully localized limit (FLL) double counting of
Ref.~\onlinecite{Anisimov}, namely $V_{dc}=U(n_d-1/2)-J/2(n_d-1)$,
where $n_d$ in the formula is computed self-consistently. We note that
this makes charge self-consistency convergence a bit more challenging,
but the results for the transition metal insulators studies here, are
almost identical to those presented above for fixed double counting.

\begin{figure}[h]
\centering{
\includegraphics[width=1\linewidth]{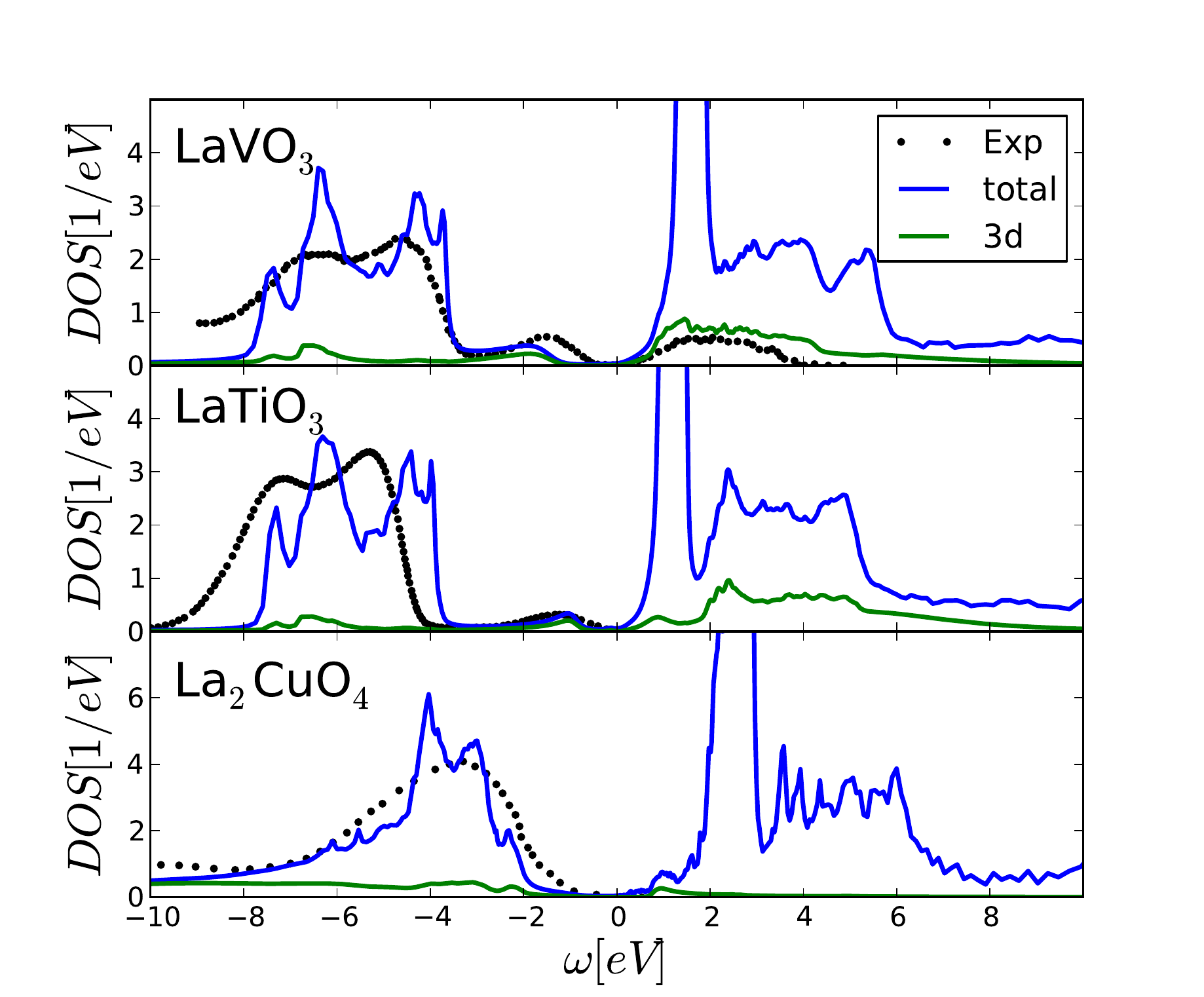}
  }
\caption{ The total DOS and its projection to the $3d$ ion within
  DFT+DMFT for insulating transition metal oxides, computed using
  standard fully localized limit double counting and fixed $U=10\,$eV
  and $J_{Hunds}=0.7\,$eV. Experimental
  photoemission is the same as in Fig.~\ref{DOSall}.  }
\label{Fig10}
\end{figure}
In Fig.~\ref{Fig10} we plot the spectral functions of all four
insulators but here computed using standard FLL
double-counting. Comparison with Fig.~\ref{DOSall} shows that the
differences between the two double-countings are extremely small and
unimportant in these cases. This is not surprising given the converged
DFT+DMFT occupancy of the correlated $d$ orbitals.
\begin {table}
\begin{center}
\begin{tabular}[c]{|l|c|c|c|}
\hline  
$n_d(DFT+DMFT)$  & fixed-DC & FLL-DC &  FLL-DC \\
                & CSC       & CSC    & non-CSC \\
\hline  
La$_2$CuO$_4$ ($R_{MT}=1.88$)&   9.031  &   9.042   &     9.049\\
\hline
LaVO$_3$      ($R_{MT}=1.97$)&   2.075  &   2.098   &     2.093\\
\hline
LaTiO$_3$     ($R_{MT}=2.01$)&   1.093  &   1.131   &     1.133\\
\hline
\end{tabular}
\caption{The occupancy of the correlated 3$d$ orbitals within
  DFT+DMFT calculation
  for insulating transition metal oxides. The first row is for
  double-counting introduced in section~\ref{Method} (fixed-DC) and full charge
  self-consistent calculation (CSC). The second row stands for fully
  localized limit double-counting (FLL-DC) and charge
  self-consistency (CSS). The last column shows $n_d$ for FLL-DC without
  charge self-consistency (non-CSS).
}
\label{Tabel1}
\end{center}
\end {table}
In Table~\ref{Tabel1} we list the 3$d$ occupancies for the
calculations presented in Fig.~\ref{DOSall} (denominated by
\textit{fixed-DC}) and for the standard FLL double counting
(denominated \textit{FLL-DC}). Note that for the early transition metal
oxides we count $t2g$ charge here, since correlations are applied to
$t2g$ set of orbitals. For La$_2$CuO$_4$, we list total 3$d$
occupancy, since the correlations are applied to all 3d electrons.  We
also tested if charge self-consistency is important for insulating
nature of these compounds. The charge-self consistent and non-charge
self-consistent calculation is denoted by CSS and non-CSS in
Table~\ref{Tabel1}. The differences in occupancies are insignificant.
The double-counting used above (fixed-DC) makes the insulating state
slightly more robust, since $n_d$ is closest to nominal valence (of the order
of $0.02$ electron less than FLL-DC), the alternative FLL
double-counting slightly increases mixed valency, while the
charge-self consistency has very small effect (of the order of
$0.005$) in these transition metal oxides.

\begin{figure}[h]
\centering{
\includegraphics[width=1\linewidth]{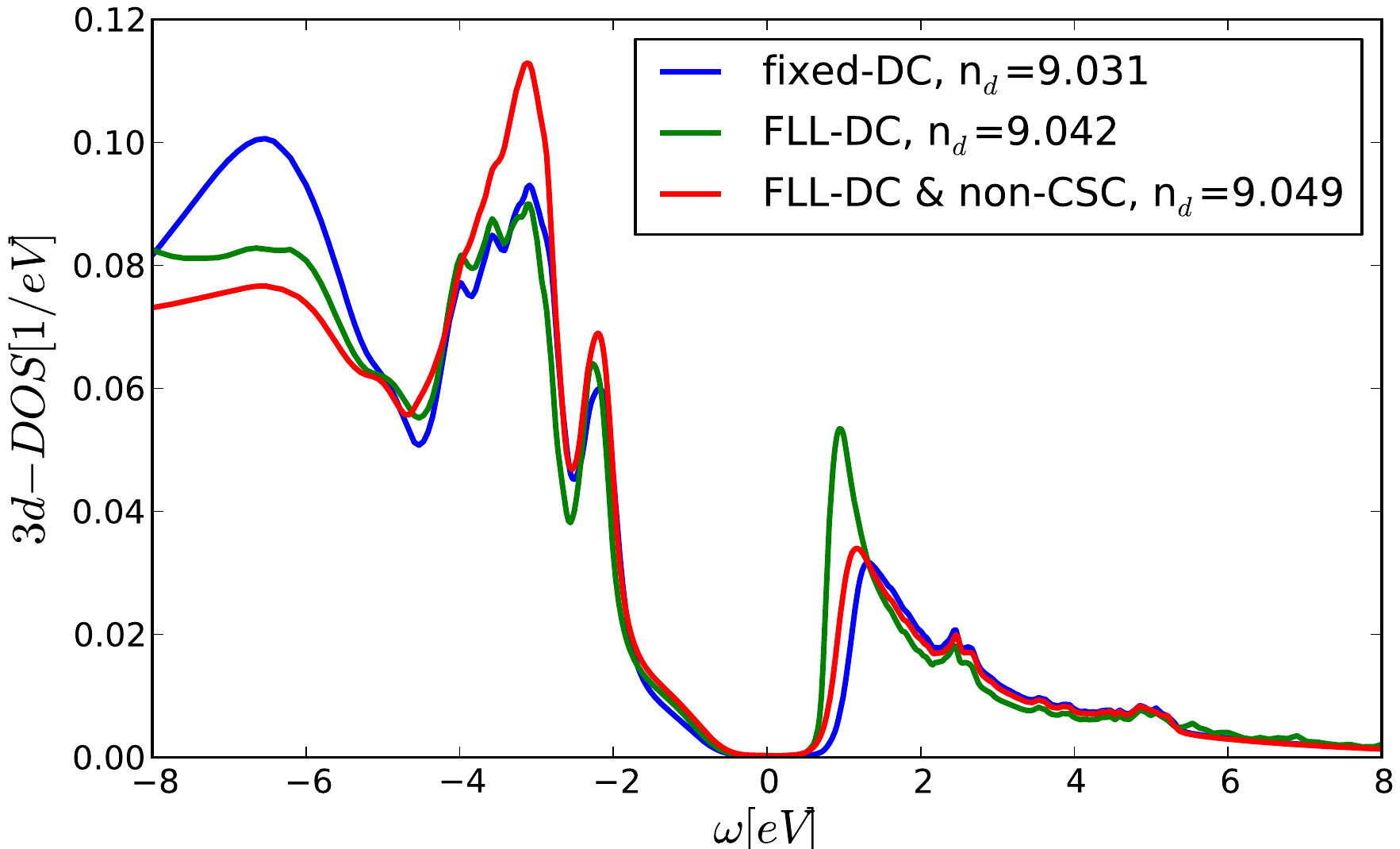}
  }
\caption{
The $3d$ density of states within DFT+DMFT for La$_2$CuO$_4$ using
the three different methodologies explained in Table~\ref{Tabel1}.
}
\label{Fig9}
\end{figure}
In Fig.~\ref{Fig9} we show the 3$d$-DOS for La$_2$CuO$_4$ in the three
types of calculations listed in table~\ref{Tabel1}. There are some minor
differences in the spectra distribution, and as expected from slightly
larger $n_d$ in FLL-DC case, the gap gets slightly smaller, but
overall these differences are very small.

The above results appear to contradict conclusions of
Ref.~\onlinecite{Chris1,Chris2}, in which the authors assert that 3$d$
occupancy of LDA+DMFT calculations are consistently too large to allow
Mott insulating state.
The authors then proposed to adjust the
double-counting energy to allow the opening of the Mott gap within
their implementation of LDA+DMFT. They traced the problem to the
$d$-occupancy being way too large even on the DFT level ($U=0$). To
define the $d$ occupancy, they constructed the maximally localized
Wannier orbitals with transition metal 3$d$ and oxygen 2-$p$ orbitals
included in the model, and excluding all the rest. In such a model,
they noticed that the 3$d$ occupancy in La$_2$CuO$_4$ is around
$n_d\approx 9.45$, in LaVO$_3$ and in LaTiO$_3$ is around
$n_{t2g}\approx 2.55$, and $n_{t2g}\approx 1.45$, respectively. These
numbers are clearly larger than numbers obtained by our method at
$U=0$ in Figs.~\ref{LVO},~\ref{LTO}, and \ref{LCO}, hence their work
revealed difficulties with the formulation and implementation of
DFT+DMFT when Wannier basis is not sufficiently localized.

\begin{figure}[h]
\centering{
\includegraphics[width=1\linewidth]{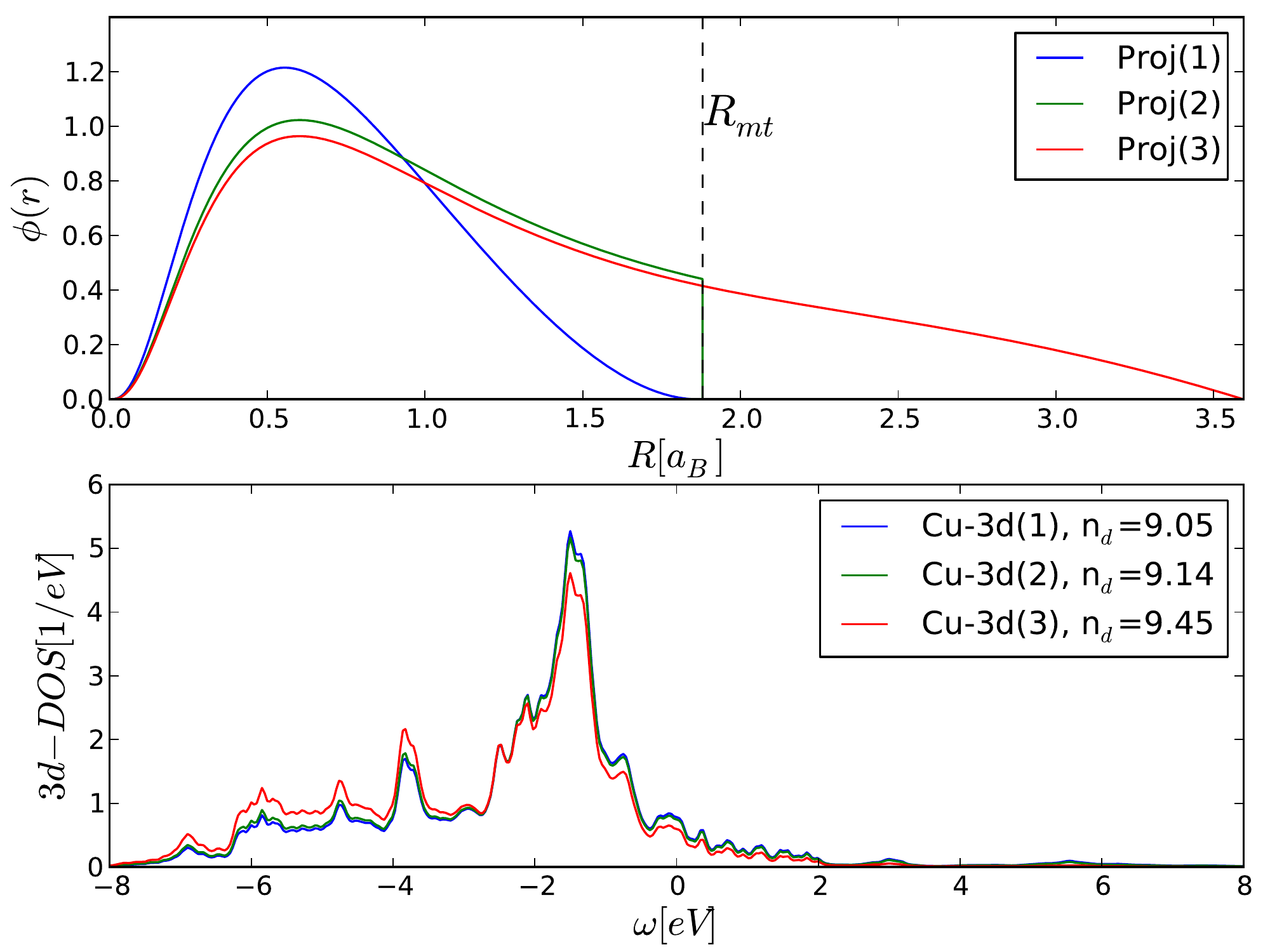}
  }
\caption{
The radial dependence of the function $\phi(r)$, which defines the
projector to the correlated $3d$ set of orbitals. Three different
projectors (Proj(1),Proj(2) and Proj(3)) are defined in the upper
panel. Lower panel shows the density of states within LDA for
La$_2$CuO$_4$ using the three different projectors defined in the
upper panel.
}
\label{Fig11}
\end{figure}

To understand the difference, we implemented a flexible real space
projectors, which generalize atomic like projectors introduced in
Ref.~\onlinecite{Haule-DMFT}, but can be extended into interstitial
region or made very localized inside MT-sphere. The form of such a
projector in real space is
\begin{eqnarray}
P(l,m,lm',\vr\vr') = Y_{lm}(\hat{\vr})\phi_l(r) \phi_l(r') Y_{lm'}(\hat{\vr}')
\end{eqnarray}
In the above calculations, the projector was constructed from
$\phi_l(r)$, which is a solution of the Dirac equation inside the
muffin-thin sphere, and is zero outside the sphere. In the actual
implementation of Ref.~\onlinecite{Haule-DMFT}, we also added the
small contribution coming from the energy derivative of the radial
wave function, which turns out not to be important in these cases.
Here we allow $\phi_l(r)$ to be any radial wave function constructed
from solution of the Dirac equation $u_l(r,E_\nu)$ with $E_\nu$
positioned at the center of the band, its first ($d
u_l(r,E_\nu)/dE_\nu\equiv \dot{u}_l$) and second derivative ($d^2
u_l(r,E_\nu)/d^2E_\nu=\ddot{u}_l$). To gain an insight into the
precise definition of the $n_d$ occupancy, we projected the Kohn-Sham
solution to the variety of projectors, spanning the range of very
localized to moderately delocalized, using a linear combination of
above defined functions: $\phi(r) = a u_l(r) + b \dot{u}_l(r) +
c\ddot{u}_l(r)$. First we choose a very localized function $\phi(r)$
entirely contained in the MT-sphere and vanishing at the
sphere with vanishing derivative. We name the corresponding projector
$Proj(1)$ in Fig.~\ref{Fig11}. (Here the MT-spheres are
chosen in such a way that the spheres of neighboring atoms
touch). Notice that such combination of $u_l$, $\dot{u}_l$ and
$\ddot{u}_l$ roughly corresponds to choosing the linearization energy at
the bottom of the corresponding bands.

The second projector $Proj(2)$ takes $u_l$ only (i.e., sets $b$ and
$c$ to zero, hence linearization energy is taken at the middle of the
corresponding band), but it truncates the radial function at the
MT-sphere. This projector is very similar to what is used in
chapter~\ref{Results} of this paper. (The precise definition, which
includes the correction due to $\dot{u}$ can be found in
Ref.\onlinecite{Haule-DMFT}.) Finally, we extended the projector to
the interstitial region with projector $Proj(3)$. Inside MT-sphere, we
choose $u_l$ as in $Proj(2)$, and outside we choose linear combination
of $u_l$, $\dot{u}_l$ and $\ddot{u}_l$ such that the function vanishes
at the nearest neighbor oxygen, and is continuous differentiable
across the MT-sphere. When projecting the Kohn-Sham solution
to the function $\phi(r)$ beyond the MT-sphere of the 3$d$-ion
$(r>R_{mt})$, we project to the plane-wave envelop functions only, to
excluded the density concentrated inside the oxygen MT-spheres, which should not be counted as transition metal charge. We
always normalize the projector, to exclude the trivial effect of
volume increase.

In Fig.~\ref{Fig11} we show the radial functions $\phi(r)$ for these
three projectors in the case of La$_2$CuO$_4$, together with the DFT
projected density of states (more precisely $-\Im(G_d(\omega))$).
As is clear from the figure, the more delocalized projector has
substantially more weight in the region between -7 to -4$\,$eV, in the
energy where oxygen is concentrated, while the localized projectors
have more weight at the upper edge of the DOS and less concentrated
around oxygen. The net result is a different occupancy $n_d$.
\begin {table}
\begin{center}
\begin{tabular}[c]{|l|c|c|c|}
\hline  
$n_d(DFT)$  & Proj(1) & Proj(2) &  Proj(3)\\
\hline  
La$_2$CuO$_4$ &   9.040  &   9.131   &     9.450\\
\hline
LaVO$_3$      &   2.096 &   2.193  & 2.872\\
\hline
LaTiO$_3$     &   1.097 &   1.195 &   1.960\\
\hline
SrVO$_3$     &   1.692 &  1.815 &  2.567\\
\hline
\end{tabular}
\end{center}
\caption{The occupancy of the 3$d$ orbitals using DFT (at $U=0$) for
  the  three projectors shown in Fig.~\ref{Fig11}.
  The MT-spheres are listed in Table~\ref{Tabel1}.
}
\label{Tabel2}
\end{table}
In table~\ref{Tabel2} we list the DFT occupancies obtained by
projecting to these three projectors and for all compounds studied
here. We again project to $t2g$ states for early transition metal
oxides, and to $eg$ and $t2g$ states in La$_2$CuO$_4$, because these
are the states which are correlated in the DMFT calculations.  For
La$_2$CuO$_4$, we notice that the two localized projectors ($Proj(1)$
and $Proj(2)$) both have occupancy close to nominal $d^9$, and that
the projector $Proj(2)$ has slightly more charge (1\% increase) than
the most localized $Proj(1)$. As shown by direct DFT+DMFT calculation
above, $Proj(2)$ gives the Mott insulating state in La$_2$CuO$_4$
irrespective of small details in double-counting (fixed-DC or FLL-DC)
or charge-self-consistency. On the other hand, $Proj(3)$, which
extends beyond the MT-boundary, contains some of the charge that
should have been assigned to other itinerant states. As a result, it
gives occupancy $n_d=9.45$, almost identical to the charge on the
Wannier orbitals of Ref.~\onlinecite{Chris1}. Since construction of
Wannier orbitals inevitable results in some fraction of electrons
being delocalized beyond the MT-boundary, it is not surprising that
the 3$d$ occupancy is similar to our more delocalized
projector. Namely, maximally localized Wannier orbitals need to
faithfully represent a set of low energy bands, hence their
localization is constrained to this condition.  We verified that
DFT+DMFT solution using $Proj(3)$ and FLL-DC results in metallic
state, similar to finding of Ref.~\onlinecite{Chris1}.

For the insulating early transition metal oxides, the $t2g$
occupancies of both localized projectors $Proj(1)$ and $Proj(2)$ are
again quite close to nominal valence, namely $d^1$ for LaTiO$_3$ and
$d^2$ for LaVO$_3$.  On the other hand, the delocalized projector
$Proj(3)$ results in much larger $n_d$, even larger than reported in
Ref.~\onlinecite{Chris1,Chris2}. Hence, Wannier orbitals in
Ref.~\onlinecite{Chris2} are more
localized than our $Proj(3)$, but less than $Proj(2)$ or
$Proj(1)$. We verified that increasing localization of the projector
always results in increased $n_d$ occupancy.
The Mott state within DFT+DMFT is again very robust using Proj(1) and
Proj(2), but not when Proj(3) is used.  Finally, the 3$d$ occupancy in
SrVO$_3$ is quite far from nominal $d^1$ valence even when using very
localized projectors, and hence this mixed-valency results in metallic
state even for very large values of local Coulomb repulsion $U$, in
agreement with experimental observation of a metallic state in SrVO$_3$.

In the early transition metal oxides, we projected the Kohn-Sham
solution to the $t2g$ states, because the center of the $eg$ states is
sufficiently above the Fermi level that it does not require dynamic
treatment within the DMFT. Since the $eg$ states strongly hybridize
with the oxygen, the $eg$ occupancy does not exactly vanish. However,
the DFT+DMFT solution is very sensitive to the correlated $t2g$
occupancy in early transition metal oxides, since the $eg$ states
behave very differently, having a large gap. Hence, the $eg$ and
$t2g$ charge should not be counted together when assessing stability
of the DMFT insulating solution, hence we presented the $t2g$ charge
only in tables~\ref{Tabel1},\ref{Tabel2}.



\section{Conclusions}

We have shown in this paper that with the DFT+DMFT methodology of
Ref.~\onlinecite{Haule-DMFT}, a reasonable qualitative agreement between
theory and experiment for the $p$ and $d$ spectra across the
transition metal series is obtained, even when the Coulomb repulsion
$U$ and $J$ are fixed across the entire series, hence no tuning
parameter is needed for qualitative description of correlated solids, which is a
requirement for any ab-initio predictive method. This was possible
because the DFT+DMFT method is implemented with a very localized
projector, where the screening of the Coulomb repulsion by other
valence states through hybridization is very efficient.

A large effort was undertaken recently by several
groups~\cite{Haule-DMFT,Amadon1,Amadon2,Lecherman,Nordstrom,Aichorn}
to implement DFT+DMFT in a way that does not require tuning
parameters, and that has \textit{ab inito} predictive power. In our
opinion, such mature state of DFT+DMFT has largely been reached, as
demonstrated on early and late transition metal oxides here.
This method gives a zeroth-order picture of the
physics in correlated materials such as transition metal
oxides. However, the position of oxygen states is not very precise
is some compounds (see LaTiO$_3$), and better treatment of exchange
would be needed to mitigate this deficiency. It was recently
proposed in Ref.\onlinecite{Held2} that additional Hartree term due
to non-local interaction $U_{pd}$ could mitigate this
problem. However, in the charge-self-consistent DFT+DMFT used here, the
Hartree terms are taken into account exactly, and therefore no extra
Hartree shifts are justified. Further corrections could come only from
better treatment of the non-local exchange.  Furthermore, the gap sizes of
Mott insulators and positions of Hubbard bands can be 
improved by calculating Coulomb $U$ more precisely from first
principles. This is an important open problem in condensed matter
theory. Methods such as GW~\cite{Kutepov} and
constrained-RPA~\cite{cRPA} show some promise in this direction, but
more work is needed to get precise enough values of $U$ for modern
DFT+DMFT codes, which use localized atomic orbitals in a large energy
window.

In our implementation of DFT+DMFT, the position of oxygen states is
not very far from its DFT value, and quite insensitive to the value of
the correlation strength, in contrast to finding of
Refs.~\onlinecite{Chris1,Chris2}.
While the position of the oxygen states in DFT are not always in very
good agreement with experiment, their small displacement does not lead
to a major failure of DFT+DMFT.
This shortcoming of LDA is known to occur in other materials (see for
example Ref.~\onlinecite{ZhipingKotliar}) and can be corrected by a
better treatment of the non-local exchange as in hybrid DFT or GW, but
not by DMFT.

We have also shown in this paper that in transition metal oxides the
self-consistent value of the correlated electronic charge $n_d$ of
LDA+DMFT is closer to nominal valence than its LDA value. A similar
finding was reported in Ref.~\onlinecite{BranchingRatios} when
studying the actinide series and its compounds. A systematic
comparison with the X-ray data confirmed that the LDA+DMFT
systematically improves the value of the correlated charge, $n_f$,
relative to its LDA value.

Finally, the DMFT method is an orbitally dependent method, and the
results depend on the choice of the correlated set of orbitals. The
convergence of the results with respect to the number of orbitals is
not possible at present, because the quantum mechanical problem
becomes too expensive to solve. The quality of the results hence rest
on the educated choice of the correlated orbital (the choice of the
projector) which determines the set of states that are treated very
precisely, by summing all local Feynman diagrams, and those that are
treated by DFT. Since DMFT truncates interaction and correlations
beyond single site, a more localized orbital is clearly a better
choice in this method. To recover similar results in a more
delocalized basis, one would clearly need to go beyond single site
approximation, which increases computational expense exponentially.

We have shown that in transition metal oxides, the $3d$ occupancies
($n_d$) on the transition metal ion are not very far from nominal
valence when sufficiently localized radial function is chosen for the
projector. This is true even on the DFT level. We have also
explicitly demonstrated that the choice of a more delocalized radial
orbital leads to valences substantially larger than the nominal
valence, which posses a problem for DMFT method, as noted in
Refs.~\onlinecite{Chris1,Chris2}. For such a choice of correlated
states, the non-local correlations would likely need to be considered
to recover similar results as in more localized case.

In conclusion, we successfully tested our implementation of the
DFT+DMFT method in $3d$ transition metal series. The method predicts
qualitative features, such as existance of a metallic or insulating
state starting from first principles. It can be made fully automated,
and hence high-throughput screening of correlated materials is an
attractive avenue for future research.

%
%
%
%
%
%
%

\section{Acknowledgments}

We thank A.~Georges, C.~H. Yee, H.~Park, and A.~J.~Millis for stimulating
discussion.  KH, GK were supported by NSF DMR--1405303, and NSF
DMR--1308141, respectively.

\end{document}